\documentclass[10pt,twoside,twocolumn,openany]{book}
\usepackage[bg-a4]{dnd}
\usepackage[english]{babel}
\usepackage[utf8]{inputenc}
\usepackage{tikz}

  \newenvironment{noindlist}
   {\begin{list}{\labelitemi}{\leftmargin=1.1em \itemindent=-.2em}}
   {\end{list}}

          \definecolor{minimise}{RGB}{255,0,0}
          \definecolor{hide}{RGB}{255,128,0}
          \definecolor{seperate}{RGB}{250,232,5}
          \definecolor{aggregate}{RGB}{0,128,192}
          \definecolor{inform}{RGB}{180,106,255}
          \definecolor{control}{RGB}{255,90,175}   
           \definecolor{enforce}{RGB}{210,0,105}       
          \definecolor{demostrate}{RGB}{0,185,0}

\newcommand{\tikzcircleminimise}[2][minimise,fill=minimise]{\tikz[baseline=-0.5ex]\draw[#1,radius=#2] (0,0) circle ;}%
\newcommand{\tikzcirclehide}[2][hide,fill=hide]{\tikz[baseline=-0.5ex]\draw[#1,radius=#2] (0,0) circle ;}%
\newcommand{\tikzcircleseperate}[2][seperate,fill=seperate]{\tikz[baseline=-0.5ex]\draw[#1,radius=#2] (0,0) circle ;}%
\newcommand{\tikzcircleagg}[2][aggregate,fill=aggregate]{\tikz[baseline=-0.5ex]\draw[#1,radius=#2] (0,0) circle ;}%
\newcommand{\tikzcircleinform}[2][inform,fill=inform]{\tikz[baseline=-0.5ex]\draw[#1,radius=#2] (0,0) circle ;}%
\newcommand{\tikzcirclecontrol}[2][control,fill=control]{\tikz[baseline=-0.5ex]\draw[#1,radius=#2] (0,0) circle ;}%
\newcommand{\tikzcircleenforce}[2][enforce,fill=enforce]{\tikz[baseline=-0.5ex]\draw[#1,radius=#2] (0,0) circle ;}%
\newcommand{\tikzcircledemo}[2][demostrate,fill=demostrate]{\tikz[baseline=-0.5ex]\draw[#1,radius=#2] (0,0) circle ;}%

\begin{document}
\fontfamily{ppl}\selectfont 



\chapter{\huge Privacy Guidelines for Internet of Things}


\section{\Large A Cheat Sheet {\footnotesize (0.1V)}}

\normalsize \noindent \textbf{How to cite:} Charith Perera (2017), \textit{Privacy Guidelines for Internet of Things: A Cheat Sheet}, Technical Report, Newcastle University, UK

\subsection{Overview}
This document presents 30 different privacy guidelines that can be used to both design and assess  IoT applications and IoT middleware platforms. These guidelines can be broadly categorised into eight categories, namely, MINIMIZE (\tikzcircleminimise{4pt}), HIDE (\tikzcirclehide{4pt}), SEPARATE (\tikzcircleseperate{4pt}), AGGREGATE (\tikzcircleagg{4pt}), INFORM (\tikzcircleinform{4pt}), CONTROL (\tikzcirclecontrol{4pt}), ENFORCE (\tikzcircleenforce{4pt}), DEMONSTRATE (\tikzcircledemo{4pt}). This document uses the following structure to introduce the each privacy guidelines. First, we describe the philosophy behind each guideline in general. Then, we present the questions that software architects need to think about when designing or assessing an IoT  platform or application. The questions slightly vary depending on whether the architect is assessing a platform or an application.


\begin{paperbox}{01. Minimise data acquisition \hfill \tikzcircleminimise{4pt} \tikzcirclehide{4pt} }
		This guideline suggests to minimise the amount of data collected or requested by an IoT platform or application. Minimisation includes:
		\begin{itemize}
		\item Minimising data types (e.g., energy consumption, water consumption, temperature)
		\item Minimum duration (e.g., hours, days, weeks, months)
		\item Minimum frequency (e.g., sampling rate: one second,  30 seconds, minutes)
		\item Minimum amount of data (e.g., kilobytes, megabytes, gigabytes)
		\end{itemize}
		The questions that  software architects need to think are:
\end{paperbox}

\begin{monsterboxnobg}{}
\noindent \textbf{Platform Assessment:} 
\begin{itemize}
\setlength{\itemsep}{0pt}
\setlength{\parskip}{0pt}
\setlength{\parsep}{0pt}  
 \item  Does the platform allows developers to control the types of data which the application receive?
 \item  Does the platform allows developers to control the duration which the application receive data?
 \item  Does the platform allows developers to control the frequency which the application receive data?
 \item  Does the platform allows developers to control the amount of data the application receive?
\end{itemize}
\noindent \textbf{Application Assessment:}

\begin{itemize}
\setlength{\itemsep}{0pt}
\setlength{\parskip}{0pt}
\setlength{\parsep}{0pt}  
\item What are the data types that need to be collected in order to achieve the tasks at hand?
\item Does the application collect more data types than necessary?
\item How long does the application need to collect data? 
\item Can the duration be  reduced  while achieving acceptable results?
\item What is the sampling frequency that need to be used to collect data in order to achieve the tasks at hand?
\item Does the application collect  data at a higher frequency than necessary? 
\item Can the data collection frequency be  reduced while achieving acceptable results?
\item Does the application gathers  more data than necessary? 
\item Can the application achieve the tasks at hand by acquiring less amount of data?
\end{itemize}
\end{monsterboxnobg}


\begin{paperbox}{02. Minimise number of data sources \hfill \tikzcircleminimise{4pt}}
This guideline suggests to minimise the number of data sources used by an IoT platform or application. Depending on the task at hand, it may be required to collect data from different sources. Multiple data sources may hold pieces of information about an individual (e.g., An activity tracking service may hold an individual's activity data while a hospital may hold his health records). Aggregation of data from multiple sources allow malicious parties to identify sensitive personal information of an individual that could lead to privacy violations.
\end{paperbox}

\begin{monsterboxnobg}{}
\noindent \textbf{Platform Assessment:} 
\begin{itemize}
\setlength{\itemsep}{0pt}
\setlength{\parskip}{0pt}
\setlength{\parsep}{0pt}  
\item Does the platform provide easy ways to manage data sources and connectivity between sources and the platform?

\item Can the platform assist the developers by pro-actively recommending alternative ways to achieve the tasks at hand instead of collecting data from large number of data sources? Detect \textit{`fall'} using a single source (e.g., wearable) vs detect \textit{`fall'} by combing three sources (e.g., wearable, smart bed, smart carpet)
 
\end{itemize}
\noindent \textbf{Application Assessment:}

\begin{itemize}
\setlength{\itemsep}{0pt}
\setlength{\parskip}{0pt}
\setlength{\parsep}{0pt}  
\item From how many data sources does this application collect data? 
\item Is it essential to collect data from all those data sources?
\end{itemize}
\end{monsterboxnobg}


\begin{paperbox}{03. Minimise raw data intake \hfill \tikzcircleminimise{4pt} \tikzcircleagg{4pt} 
}
Whenever possible, IoT applications should reduce the amount of raw\footnote{Unprocessed and un-fused data can be identified as Raw data (also called \textit{primary context}). For example, X-axis value of an accelerometer can be identified as raw data. Knowledge (e.g. current activity = `walking') generated by processing and fusing X-, Y-, and Z-axis values together can be identified as process data (also called \textit{secondary context}).} data intake. Raw data  could lead to secondary usage and privacy violation. Therefore, IoT platforms should consider converting (or transforming) raw data into secondary context data. For example, IoT applications can extract orientation (e.g. sitting, standing, walking) by processing  accelerometer data and  store only the results (i.e. secondary context) and delete the raw accelerometer data.
\end{paperbox}

\begin{monsterboxnobg}{}
\noindent \textbf{Platform Assessment:} 
\begin{itemize}
\setlength{\itemsep}{0pt}
\setlength{\parskip}{0pt}
\setlength{\parsep}{0pt}  
 \item  Does the platform allows developers to transform raw data (primary context)  into abstract secondary derived data (secondary context)?
 \item How easy it is to develop an extension that transforms  raw data into abstract secondary derived data?
 \item Does the platform actively discourage developers from accepting raw data into the system?
\end{itemize}
\noindent \textbf{Application Assessment:}

\begin{itemize}
\setlength{\itemsep}{0pt}
\setlength{\parskip}{0pt}
\setlength{\parsep}{0pt}  
\item How much raw data does the application acquires?
\item Does the application acquire any raw data that can be transformed into secondary data without affecting the overall outcome of the system?

\end{itemize}
\end{monsterboxnobg}


\begin{paperbox}{04. Minimise knowledge discovery \hfill \tikzcircleminimise{4pt}}
This guideline suggests to minimise the amount of knowledge discovered within an IoT application. IoT applications should only discover the knowledge  necessary to achieve their  primary objectives. For example, if the objective  is to recommend food plans, it should not attempt to infer users' health status without their explicit permission.
\end{paperbox}

\begin{monsterboxnobg}{}
\noindent \textbf{Platform Assessment:} 
\begin{itemize}
\setlength{\itemsep}{0pt}
\setlength{\parskip}{0pt}
\setlength{\parsep}{0pt}  
 \item How easy it is to develop an extension to discover secondary knowledge by processing raw data?
 \item Does the platform keeps track of how much knowledge is discovered over time?  
 \item Does the platform provide recommendation to  developers on achieving the tasks at hand without discovering knowledge more than necessary?
\end{itemize}
\noindent \textbf{Application Assessment:}

\begin{itemize}
\setlength{\itemsep}{0pt}
\setlength{\parskip}{0pt}
\setlength{\parsep}{0pt}  
\item Does the application produce more secondary knowledge than it required to achieve the tasks at hand? 
\end{itemize}
\end{monsterboxnobg}


\begin{paperbox}{05. Minimise data storage \hfill \tikzcircleminimise{4pt}}
This guideline suggests to minimise the amount of data (i.e. primary or secondary) stored by an IoT application. Any  piece  of  data  that  is  not required to perform a certain task should  be deleted. For example, raw data can be deleted once secondary contexts are derived. Further, personally identifiable data may  be deleted without storing.
\end{paperbox}

\begin{monsterboxnobg}{}
\noindent \textbf{Platform Assessment:} 
\begin{itemize}
\setlength{\itemsep}{0pt}
\setlength{\parskip}{0pt}
\setlength{\parsep}{0pt}  
\item Can the platform be configured to  delete raw data after deriving essential secondary knowledge?
 \item Can the platform identify unnecessary data storage and notify the developers?
  \item Can the platform identify unnecessary data storage and delete the autonomously? 
   \item Can the platform meter and monitor data storage?

\end{itemize}

\noindent \textbf{Application Assessment:}

\begin{itemize}
\setlength{\itemsep}{0pt}
\setlength{\parskip}{0pt}
\setlength{\parsep}{0pt}  
\item  Does the application  delete raw data after deriving essential secondary knowledge?
\end{itemize}
\end{monsterboxnobg}


\begin{paperbox}{06. Minimise data retention period \hfill \tikzcircleminimise{4pt} \tikzcirclehide{4pt}}
This guideline suggests to minimise the duration for which data is stored (i.e. avoid retaining data for longer than needed).  Long retention periods provide more time for malicious parties to attempt accessing the data in unauthorized manner. Privacy risks are also increased because long retention periods could lead to unconsented secondary usage.
\end{paperbox}

\begin{monsterboxnobg}{}
\noindent \textbf{Platform Assessment:} 
\begin{itemize}
\setlength{\itemsep}{0pt}
\setlength{\parskip}{0pt}
\setlength{\parsep}{0pt}  
\item Can the platform be configured to  delete  data on pre-defined set time / set frequency / events?
\item Can the platform notify developers on optimum data removal? 
\end{itemize}

\noindent \textbf{Application Assessment:}

\begin{itemize}
\setlength{\itemsep}{0pt}
\setlength{\parskip}{0pt}
\setlength{\parsep}{0pt}  
\item  Does the application  retain data longer time periods than  necessary ?
\end{itemize}
\end{monsterboxnobg}


\begin{paperbox}{07. Hidden data routing \hfill \tikzcirclehide
{4pt}}
In IoT, data is generated within sensing devices. The data analysis typically happens within cloud servers. Therefore, data is expected to travel between different types of computational nodes before arriving at the processing cloud servers. This type of routing could reveal user locations and usage from anyone conducting network surveillance or traffic analysis. To make it more difficult for Internet activities to be traced back to the users, this guideline suggests that IoT applications should support and employ anonymous routing mechanism (e.g., torproject.org). 
\end{paperbox}

\begin{monsterboxnobg}{}
\noindent \textbf{Platform Assessment:} 
\begin{itemize}
\setlength{\itemsep}{0pt}
\setlength{\parskip}{0pt}
\setlength{\parsep}{0pt}  
\item Does the platform provides built in support for anonymous routing mechanism?
\item If so, can anonymous routing mechanism be enabled easily?
\item Does the platform reminds the developers, if the application developed on top of the platform can effectively use anonymous routing mechanism?
 
\end{itemize}

\noindent \textbf{Application Assessment:}

\begin{itemize}
\setlength{\itemsep}{0pt}
\setlength{\parskip}{0pt}
\setlength{\parsep}{0pt}  
\item Can the application use anonymous routing mechanisms?
\item If so, does the application use anonymous routing mechanisms?
\end{itemize}
\end{monsterboxnobg}


\begin{paperbox}{08. Data anonymisation \hfill \tikzcirclehide
{4pt}}
This guideline suggests to remove personally identifiable information before the data gets used by IoT applications  so that the people described by the data remain anonymous. Removal of personally identifiable information  reduces the  risk of unintended disclosure and privacy violations. 
\end{paperbox}

\begin{monsterboxnobg}{}
\noindent \textbf{Platform Assessment:} 
\begin{itemize}
\setlength{\itemsep}{0pt}
\setlength{\parskip}{0pt}
\setlength{\parsep}{0pt}  
\item Does the platform provide easy way to remove personally identifiable information?
\item  Does the platform provide variety of different anonymisation techniques?
\item Can the platform autonomously recognise personally identifiable information?
 
\end{itemize}

\noindent \textbf{Application Assessment:}

\begin{itemize}
\setlength{\itemsep}{0pt}
\setlength{\parskip}{0pt}
\setlength{\parsep}{0pt}  
\item Does the application deal with personally identifiable information at all?
\item Is it possible to achieve the tasks at hand without dealing with  personally identifiable information at all?
\item What personally identifiable information does the application collects? 
\item Is it possible to anonymise the data collected and still achieve the tasks at hand?
\end{itemize}
\end{monsterboxnobg}


\begin{paperbox}{09. Encrypted data communication \hfill \tikzcirclehide
{4pt}}
This guideline suggests that different components in an IoT application should consider encrypted data communication wherever possible. Encrypted data communication would reduce the potential privacy risks due to unauthorised access during data transfer between components. There are multiple data communication approaches based on the components involved in an IoT application, namely, 1) device-to-device, 2) device-to-gateway, 3) device-to-cloud, and 4) gateway-to-cloud. Sensor data communication can be encrypted using symmetric encryption AES 256 in the application layer. Typically, device-to-device communications are encrypted at the link layer using special electronic hardware included in the radio modules. Gateway-to-cloud communication is typically secured through HTTPS using  Secure Sockets Layer (SSL) or Transport Layer Security (TLS).
\end{paperbox}

\begin{monsterboxnobg}{}

\noindent \textbf{Platform Assessment:} 
\begin{itemize}
\setlength{\itemsep}{0pt}
\setlength{\parskip}{0pt}
\setlength{\parsep}{0pt}  
\item Does the platform provide easy way to encrypt data communications?
\item Does the platform provide variety of different encryption techniques?
\item Does the platform encrypt data by default when communicating? 
 
\end{itemize}

\noindent \textbf{Application Assessment:}

\begin{itemize}
\setlength{\itemsep}{0pt}
\setlength{\parskip}{0pt}
\setlength{\parsep}{0pt}  
\item Does the application encrypts its data communications? If not, why?
\item Which encryption technique does the application employs?
\end{itemize}
\end{monsterboxnobg}

\begin{paperbox}{10. Encrypted data processing  \hfill \tikzcirclehide
{4pt}}
This guideline suggests to process data while encrypted.  Encryption is the process of encoding data in such a way that only authorised parties can read it. However, sometimes, the party who is responsible for processing data should not be allowed to read data. In such circumstances, it is important to process data while they are in encrypted form. For example, homomorphic encryption  is a form of encryption that allows computations to be carried out on cipher-text, thus generating an encrypted result which, when decrypted, matches the result of operations performed on the  plain-text.
\end{paperbox}
\begin{monsterboxnobg}{}
\noindent \textbf{Platform Assessment:} 
\begin{itemize}
\setlength{\itemsep}{0pt}
\setlength{\parskip}{0pt}
\setlength{\parsep}{0pt} 
\item Can the platform process data while encrypted?  
\item Does the platform provide easy way to add new encrypted data processing techniques?

\end{itemize}

\noindent \textbf{Application Assessment:}

\begin{itemize}
\setlength{\itemsep}{0pt}
\setlength{\parskip}{0pt}
\setlength{\parsep}{0pt}  
\item Does the application process data while encrypted?  

\end{itemize}
\end{monsterboxnobg}


\begin{paperbox}{11. Encrypted data storage \hfill \tikzcirclehide
{4pt}}
This guideline suggests that IoT applications should store data in encrypted form. Encrypted data storage reduces any privacy violations due to   malicious  attacks and unauthorised access. Data encryption can be applied in different levels from sensors  to the cloud. Depending on the circumstances, data can be encrypted using both hardware and software technologies.
\end{paperbox}

\begin{monsterboxnobg}{}
\noindent \textbf{Platform Assessment:} 
\begin{itemize}
\setlength{\itemsep}{0pt}
\setlength{\parskip}{0pt}
\setlength{\parsep}{0pt}  
\item Does the platform provide easy way to encrypt data storage?
\item Does the platform provide variety of different encryption techniques?
\item Does the platform encrypt data by default when storing? 
 
\end{itemize}

\noindent \textbf{Application Assessment:}

\begin{itemize}
\setlength{\itemsep}{0pt}
\setlength{\parskip}{0pt}
\setlength{\parsep}{0pt}  
\item Does the application encrypts its stored data? If not, why?
\item Which encryption technique does the application employs?
\end{itemize}
\end{monsterboxnobg}


\begin{paperbox}{12. Reduce data granularity \hfill \tikzcirclehide
{4pt}}
The granularity is the level of depth represented by the data. High granularity refers to  atomic grade of detail and low granularity zooms out into a summary view of data. For example, dissemination of location can be considered as coarse-grained and full address can be considered as fine-grained. Therefore, releasing fine grained information always has more privacy risks than coarse-grained data as they contain more information. Data granularity has a direct impact on the quality of the data as well as the accuracy of the results produced by processing such data. IoT applications should request the minimum level of granularity that is required to perform their primary tasks. Higher level of granularity could lead to secondary data usage and eventually privacy violations.
\end{paperbox}

\begin{monsterboxnobg}{}
\noindent \textbf{Platform Assessment:} 
\begin{itemize}
\setlength{\itemsep}{0pt}
\setlength{\parskip}{0pt}
\setlength{\parsep}{0pt}  
\item Does the platform autonomously identifies data columns where data granularity reductions can be applied?
\item Does the platform provide recommendations on  how-to and when-to apply data granularity reductions techniques?
\item What kind of data granularity reduction techniques does the platform provides?
 
\end{itemize}

\noindent \textbf{Application Assessment:}

\begin{itemize}
\setlength{\itemsep}{0pt}
\setlength{\parskip}{0pt}
\setlength{\parsep}{0pt}  
\item Does the application reduces data granularity when ever it does not impact the objectives?
\end{itemize}
\end{monsterboxnobg}


\begin{paperbox}{13. Query answering \hfill \tikzcircleminimise{4pt} \tikzcirclehide {4pt}}
This guideline suggests to release high-level answers  to the query when dissemination without releasing raw data. For example, a sample query would be \textit{`how energy efficient a particular household is?'} where the answer would be in 0-5 scale. Raw data can always lead to privacy violations due to secondary usage. One such implementation is openPDS/SafeAnswers   where it allows users to collect, store, and give high level answers to the queries while protecting their privacy.
\end{paperbox}

\begin{monsterboxnobg}{}
\noindent \textbf{Platform Assessment:} 
\begin{itemize}
\setlength{\itemsep}{0pt}
\setlength{\parskip}{0pt}
\setlength{\parsep}{0pt}  
\item Does the platform allow applications to share data with external entities?
\item Does the platform provide query-able interface that meters, monitors, and filters out going data?
 
\end{itemize}

\noindent \textbf{Application Assessment:}

\begin{itemize}
\setlength{\itemsep}{0pt}
\setlength{\parskip}{0pt}
\setlength{\parsep}{0pt}  
\item Does the application share data with external entities? 
\item Is it essential to share data with external entities?
\item Is it possible to share data with Q-A pattern without sharing raw data?
\item Does the application provide query-able interface that meters, monitors, and filters out going data?

\end{itemize}
\end{monsterboxnobg}


\begin{paperbox}{14. Repeated query blocking \hfill \tikzcircleminimise{4pt} \tikzcirclehide{4pt}}
This guideline goes hand-in-hand with the {Query answering} guideline. When answering queries, IoT applications need to make sure that they block any malicious attempts to discover knowledge that violates user privacy through repeated queries (e.g, analysing intersections of multiple results).
\end{paperbox}

\begin{monsterboxnobg}{}
\noindent \textbf{Platform Assessment:} 
\begin{itemize}
\setlength{\itemsep}{0pt}
\setlength{\parskip}{0pt}
\setlength{\parsep}{0pt}  
\item Does the platform provide query-able interface that meters, monitors, and filters incoming queries in order to make sure malicious repeated are blocked?
 
\end{itemize}

\noindent \textbf{Application Assessment:}

\begin{itemize}
\setlength{\itemsep}{0pt}
\setlength{\parskip}{0pt}
\setlength{\parsep}{0pt}  
\item Does the application provide query-able interface that meters, monitors, and filters incoming queries in order to make sure malicious repeated are blocked? 
\end{itemize}
\end{monsterboxnobg}


\begin{paperbox}{15. Distributed  data processing \hfill \tikzcircleseperate{4pt}}
This guideline suggests that an IoT application should process data in a distributed manner. Similar, approaches are widely used in traditional wireless sensor network domain. Distributed processing avoids  centralised large-scale data gathering. As a result, it deters any  unauthorised data access attempts.
\end{paperbox}

\begin{monsterboxnobg}{}
\noindent \textbf{Platform Assessment:} 
\begin{itemize}
\setlength{\itemsep}{0pt}
\setlength{\parskip}{0pt}
\setlength{\parsep}{0pt}  
\item Does the platform supports distributed processing?
\item If so, does the platform support different types of  nodes that varies in computational capabilities? 
\end{itemize}

\noindent \textbf{Application Assessment:}

\begin{itemize}
\setlength{\itemsep}{0pt}
\setlength{\parskip}{0pt}
\setlength{\parsep}{0pt}  
\item Can the application process data in a distributed fashion without affecting the primary objectives?
\item Does the application  process data in a distributed fashion?
\end{itemize}
\end{monsterboxnobg}


\begin{paperbox}{16. Distributed data storage \hfill \tikzcircleseperate{4pt}}
This guideline recommends storing data in a distributed manner. Distributed  data storage  reduces any privacy violation due to   malicious  attacks and unauthorised access.  It  also reduces privacy risks due to unconsented secondary knowledge discovery.
\end{paperbox}

\begin{monsterboxnobg}{}
\noindent \textbf{Platform Assessment:} 
\begin{itemize}
\setlength{\itemsep}{0pt}
\setlength{\parskip}{0pt}
\setlength{\parsep}{0pt}  
\item Does the platform supports distributed storage?
\item If so, does the platform support different types of  nodes that varies in computational capabilities?  
 
\end{itemize}

\noindent \textbf{Application Assessment:}

\begin{itemize}
\setlength{\itemsep}{0pt}
\setlength{\parskip}{0pt}
\setlength{\parsep}{0pt}  
\item Can the application store data in a distributed fashion without affecting the primary objectives?
\item Does the application  storage data in a distributed fashion? 
\end{itemize}
\end{monsterboxnobg}


\begin{paperbox}{17. Knowledge discovery based aggregation \hfill \tikzcircleagg{4pt}}
Aggregation of information over groups of attributes or groups of individuals, restricts the amount of detail in the personal data that remains. This guideline suggests to discover knowledge  though aggregation and replace raw data with discovered new knowledge. For example, \textit{`majority  of people who visited the park on [particular date] were young students'} is an aggregated result that is sufficient (once collected over a time period)  to perform further time series based sales performance analysis of a near-by shop. Exact timings of the crowd movements are not necessary to achieve this objective.

\end{paperbox}

\begin{monsterboxnobg}{}
\noindent \textbf{Platform Assessment:} 
\begin{itemize}
\setlength{\itemsep}{0pt}
\setlength{\parskip}{0pt}
\setlength{\parsep}{0pt}  
\item Does the platform provides an easy ways to monitor and track new knowledge discovery?
 \item Does the platform recommend developers on how to perform knowledge discovery to reduce potential privacy violations and risks?
\end{itemize}

\noindent \textbf{Application Assessment:}

\begin{itemize}
\setlength{\itemsep}{0pt}
\setlength{\parskip}{0pt}
\setlength{\parsep}{0pt}  
\item Does the application discover new knowledge by aggregating different types of data?
\item If so, does the aggregations make it difficult to derive personally identifiable information?

\end{itemize}
\end{monsterboxnobg}


\begin{paperbox}{18. Geography based aggregation \hfill \tikzcircleagg{4pt}}
This guideline recommends to aggregate data using geographical boundaries. For example, a query would be \textit{`how many electric vehicles used in each city in UK'}. The results to this query would be an aggregated number unique to the each city. It is not required to collect or store detailed about individual electric vehicle.
\end{paperbox}

\begin{monsterboxnobg}{}
\noindent \textbf{Platform Assessment:} 
\begin{itemize}
\setlength{\itemsep}{0pt}
\setlength{\parskip}{0pt}
\setlength{\parsep}{0pt}  
\item Does the platform supports geography based aggregation?
\item Does the platform provides easy to use geo-marking tools to support geography based aggregation?
\item What kinds of geo marking techniques are supported? visual, cities, states, countries, counties, etc.?
 
\end{itemize}

\noindent \textbf{Application Assessment:}

\begin{itemize}
\setlength{\itemsep}{0pt}
\setlength{\parskip}{0pt}
\setlength{\parsep}{0pt}  
\item Can the application use geography based aggregation without affecting the results?
\item If so, does the application use geography based aggregation? 
\end{itemize}
\end{monsterboxnobg}


\begin{paperbox}{19. Chain aggregation \hfill \tikzcircleagg{4pt}}
This guideline suggests to perform aggregation on-the-go while moving data from one node to another. For example, if the query requires a count or average, it can be done without pulling all the data items to a centralised location. Data will be sent from one node to another until all the nodes get a chance to respond.  Similar techniques are successfully used in wireless sensor networks. This type of technique reduces the amount of data gathered by a centralised node (e.g. cloud server). Further, such aggregation also eliminates raw data from the results by reducing the risk of secondary data usage.
\end{paperbox}

\begin{monsterboxnobg}{}
\noindent \textbf{Platform Assessment:} 
\begin{itemize}
\setlength{\itemsep}{0pt}
\setlength{\parskip}{0pt}
\setlength{\parsep}{0pt}  
\item Does the platform supports chain based aggregation? 
\item How the platforms create chains? algorithms? considered criteria?
 
\end{itemize}

\noindent \textbf{Application Assessment:}

\begin{itemize}
\setlength{\itemsep}{0pt}
\setlength{\parskip}{0pt}
\setlength{\parsep}{0pt}  
\item Can the application use chain based aggregation without affecting the results?
\item If so, does the application use chain based aggregation? 
\end{itemize}
\end{monsterboxnobg}


\begin{paperbox}{20. Time-Period based aggregation \hfill \tikzcircleagg{4pt}}
This guideline suggests to aggregate data over  time  (e.g. days, week, months). Aggregation reduces the granularity of data and also reduces the secondary usage that could lead to privacy violations. For example, energy consumption of a given house can be acquired and represented in aggregated form as 160 kWh per month instead of gathering energy consumption on daily or hourly basis.
\end{paperbox}

\begin{monsterboxnobg}{}
\noindent \textbf{Platform Assessment:} 
\begin{itemize}
\setlength{\itemsep}{0pt}
\setlength{\parskip}{0pt}
\setlength{\parsep}{0pt}  
\item Does the platform supports time based aggregation?  
 \item What kinds of time based aggregations does the platform support? second, minutes, hours, days, weeks, months, years?
\end{itemize}

\noindent \textbf{Application Assessment:}

\begin{itemize}
\setlength{\itemsep}{0pt}
\setlength{\parskip}{0pt}
\setlength{\parsep}{0pt}  
\item Can the application use time based aggregation without affecting the results?
\item If so, does the application use time based aggregation? 
\end{itemize}
\end{monsterboxnobg}


\begin{paperbox}{21. Category based aggregation \hfill \tikzcircleagg{4pt}}
Categorisation based aggregation approaches can be used to reduce the granularity of the raw data. For example, instead of using exact value (e.g. 160 kWh per month), energy consumption of a given house can be represented as 150-200 kWh per month. Time-Period based and category based aggregation can be combined together to reduce data granularity.
\end{paperbox}

\begin{monsterboxnobg}{}
\noindent \textbf{Platform Assessment:} 
\begin{itemize}
\setlength{\itemsep}{0pt}
\setlength{\parskip}{0pt}
\setlength{\parsep}{0pt}  
\item Does the platform supports category based aggregation?  
 \item What kinds of category based aggregations does the platform support? age (e.g., 20-25 years), hear rate (e.g., low, normal, high)?
  \item Can the platform automatically recommend ideal categorization approach? e.g., based on knowledge, based on past experience?
\end{itemize}

\noindent \textbf{Application Assessment:}

\begin{itemize}
\setlength{\itemsep}{0pt}
\setlength{\parskip}{0pt}
\setlength{\parsep}{0pt}  
\item Can the application use category based aggregation without affecting the results?
\item If so, does the application use category based aggregation? 
\end{itemize}
\end{monsterboxnobg}


\begin{paperbox}{22. Information Disclosure \hfill \tikzcircleinform{4pt} \tikzcircledemo{4pt}}
This  guideline suggests that data subjects should be adequately informed whenever data they own is acquired, processed, and disseminated. Inform can take place at any stage of the data life cycle. Further, inform can be broadly divided into two categories: pre-inform and   post-inform. Pre-inform takes place before data enters to a given data life cycle  phase. Post-inform takes place soon after data leaves a given data life cycle  phase.
\end{paperbox}

\begin{monsterboxnobg}{}
\begin{noindlist}
\item  \textit{Consent and Data Acquisition:} what is the purpose of the data acquisition?, What types of data are requested?, What is the level of granularity?, What are the rights of the data subjects? 

\item  \textit{Data Pre-Processing:} what data will be taken into the platform?, what data will be thrown out?, what kind of pre-processing technique will be employed?, what are the purposes of pre-processing data?, what techniques will be used to protect user privacy?

\item  \textit{Data Processing and Analysis:}  what type of data will be analysed?, what knowledge will be discovered?, what techniques will be used?.

\item  \textit{Data Storage:} what data items will be stored?
 how long they will be stored? what technologies are used to store data (e.g. encryption techniques)? is it centralised or distributed storage?  will there be any back up processes?

\item \textit{Data Dissemination:} with whom the data will be shared? what rights will receivers have? what rights will data subjects have?
\end{noindlist}

\noindent \textbf{Platform Assessment:} 
\begin{itemize}
\setlength{\itemsep}{0pt}
\setlength{\parskip}{0pt}
\setlength{\parsep}{0pt}  
\item What are the techniques does the platform use to reach the data owners? SMS, emails, notifications (virtual / physical), etc.
 \item Can the platform be preconfigured to disclose information at certain events or times?
\end{itemize}

\noindent \textbf{Application Assessment:}

\begin{itemize}
\setlength{\itemsep}{0pt}
\setlength{\parskip}{0pt}
\setlength{\parsep}{0pt}  
\item Does the application disclose information to the data owners?
\item What techniques are used reach the data owners?
\item How frequently does the application disclose information?
\end{itemize}
\end{monsterboxnobg}


\begin{paperbox}{23. Control \hfill \tikzcirclecontrol{4pt} \tikzcircleenforce{4pt}}
This guideline recommends providing privacy control mechanisms for data subjects. Control mechanisms will allow data owners to manage data based on their preference. There are different aspects that the data owner may like to control.  However, controlling is a time consuming task and not every data owner will have the expertise to make such decisions.

Therefore, it is a software  architect's responsibility to carefully go through the following list of possibilities and determine what kind of controls are useful and relevant to data owners in a given IoT  application context. Further, it is important to provide some kind of default set of options for data owners to choose from, specially in the cases where data subjects do not have sufficient knowledge. Some   potential aspects that a data owner may like to control are 1) data granularity, 2) anonymisation technique, 3) data retention period, 4) data dissemination.
\end{paperbox}

\begin{monsterboxnobg}{}
\noindent \textbf{Platform Assessment:} 
\begin{itemize}
\setlength{\itemsep}{0pt}
\setlength{\parskip}{0pt}
\setlength{\parsep}{0pt}  
\item What kinds of privacy controls does the platform can offer to the end users? (e.g.,  data granularity, anonymisation technique, data retention period, data dissemination)
\item Can the platform make recommendations on which controls to be provided to the end users based on the application? 
 
\end{itemize}

\noindent \textbf{Application Assessment:}

\begin{itemize}
\setlength{\itemsep}{0pt}
\setlength{\parskip}{0pt}
\setlength{\parsep}{0pt}  
\item What kinds of privacy controls does the application provides to the end users? (e.g.,  data granularity, anonymisation technique, data retention period, data dissemination)
\end{itemize}
\end{monsterboxnobg}


\begin{paperbox}{24. Logging \hfill \tikzcircledemo{4pt}}
This guideline suggests to log events during all  phases.  It  allows  both internal and external parties to examine what has happened in the past  to  make  sure a given system has performed as promised. Logging  could include but not limited to event traces, performance parameters, timestamps, sequences of operations performed over data, any human interventions.  For example, a log may include the timestamps of data arrival, operations performed in order to anonymise data, aggregation techniques performed, and so on.
\end{paperbox}

\begin{monsterboxnobg}{}
\noindent \textbf{Platform Assessment:} 
\begin{itemize}
\setlength{\itemsep}{0pt}
\setlength{\parskip}{0pt}
\setlength{\parsep}{0pt}  
\item What kind of information can be logged?
\item How the logs are managed, access controlled, reliability, traceability?
\item Are the logs reliable and trustworthy? 
\item What tools are provided to analyse and interpret logs?
 
\end{itemize}

\noindent \textbf{Application Assessment:}

\begin{itemize}
\setlength{\itemsep}{0pt}
\setlength{\parskip}{0pt}
\setlength{\parsep}{0pt}  
\item What kind of information are be logged?
\item How the logs are managed, access controlled, reliability, traceability?
\item Are the logs reliable and trustworthy? 
\item What tools are provided to analyse and interpret logs?
\end{itemize}
\end{monsterboxnobg}


\begin{paperbox}{25. Auditing \hfill \tikzcircledemo{4pt}}
This guideline suggests to perform  systematic and independent examinations of logs, procedures, processes, hardware and software specifications, and so on. The logs  above could play a significant role in this processes. Non-disclosure agreements may be helpful to allow auditing some parts of the classified data analytics processes.
\end{paperbox}

\begin{monsterboxnobg}{}
\noindent \textbf{Platform Assessment:} 
\begin{itemize}
\setlength{\itemsep}{0pt}
\setlength{\parskip}{0pt}
\setlength{\parsep}{0pt}  
\item Does the application get audited? how often?
\item Who conducts the audit? are they reliable and trustworthy?
\item What components get audited?
 
\end{itemize}

\noindent \textbf{Application Assessment:}

\begin{itemize}
\setlength{\itemsep}{0pt}
\setlength{\parskip}{0pt}
\setlength{\parsep}{0pt}  
\item Does the application get audited? how often?
\item Who conducts the audit? are they reliable and trustworthy?
\item What components get audited?

\end{itemize}
\end{monsterboxnobg}


\begin{paperbox}{26. Open Source \hfill \tikzcircledemo{4pt}}
Making source code of an IoT application open allows any external party to review code. Such reviews can be used as a form of  compliance demonstration. This allows external parties to examine the code bases to verify and determine whether a given application or platform has taken all measures to protect user privacy.
\end{paperbox}

\begin{monsterboxnobg}{}
\noindent \textbf{Platform Assessment:} 
\begin{itemize}
\setlength{\itemsep}{0pt}
\setlength{\parskip}{0pt}
\setlength{\parsep}{0pt}  
\item Is the platform open sourced? well maintained and documented?
\item How many external parties seems to work with the code?
\item Is there any formal backing from major open source initiatives (e.g. Apache, Eclipse) 
\item Does this platform use any well known open source frameworks to develop certain functionalities?
\end{itemize}

\noindent \textbf{Application Assessment:}

\begin{itemize}
\setlength{\itemsep}{0pt}
\setlength{\parskip}{0pt}
\setlength{\parsep}{0pt}  
\item  Is the platform open sourced?  well maintained and documented?
\item How many external parties seems to work with the code?
\item Does this application use any well known open source frameworks to develop certain functionalities?
\end{itemize}
\end{monsterboxnobg}


\begin{paperbox}{27. Data Flow \hfill \tikzcircledemo{4pt}}
Data flow diagrams (e.g. Data Flow Diagrams  used by Unified Modelling Language) allow interested parties to understand how data flows within a given  IoT application and how data is being treated. Therefore, DFDs can be used as a form of a compliance demonstration. 
\end{paperbox}

\begin{monsterboxnobg}{}
\noindent \textbf{Platform Assessment:} 
\begin{itemize}
\setlength{\itemsep}{0pt}
\setlength{\parskip}{0pt}
\setlength{\parsep}{0pt}  
\item Does the platform visualise what happen to data within the platform? (e.g., service compositions and work flows)
 
\end{itemize}

\noindent \textbf{Application Assessment:}

\begin{itemize}
\setlength{\itemsep}{0pt}
\setlength{\parskip}{0pt}
\setlength{\parsep}{0pt}  
\item Does the DFD diagrams of the IoT application available?
\end{itemize}
\end{monsterboxnobg}


\begin{paperbox}{28. Certification \hfill \tikzcircledemo{4pt}}
In this context, certification refers to the confirmation of certain characteristics of an system and process. Typically, certifications are given by a neutral authority. Certification will add  trustworthiness to IoT applications. TRUSTe (truste.com) Privacy Seal  is one example,  even though none of the existing certifications are  explicitly designed to certify IoT   applications.
\end{paperbox}
\begin{monsterboxnobg}{}
\noindent \textbf{Platform Assessment:} 
\begin{itemize}
\setlength{\itemsep}{0pt}
\setlength{\parskip}{0pt}
\setlength{\parsep}{0pt}  
\item Does the platform have any privacy or security related certification?
 \item Does the platform have received any rating or award related to privacy or security??
\end{itemize}

\noindent \textbf{Application Assessment:}

\begin{itemize}
\setlength{\itemsep}{0pt}
\setlength{\parskip}{0pt}
\setlength{\parsep}{0pt}  
\item Does the application have acquired any privacy or security related certification?
 \item Does the application have received any rating or award related to privacy or security??
\end{itemize}
\end{monsterboxnobg}


\begin{paperbox}{29. Standardisation \hfill \tikzcircledemo{4pt}}
This guideline suggests to follow standard practices as a way to demonstrate privacy protection capabilities. Industry wide standards (e.g. AllJoyn allseenalliance.org) typically inherit security measures that would reduce some privacy risks as well. This refers to the process of implementing and developing technical standards. Standardisation can help to maximise compatibility, interoperability, safety, repeatability, or quality. Standardisation will help external parties to easily understand the inner workings of a given IoT application.
\end{paperbox}
\begin{monsterboxnobg}{}
\noindent \textbf{Platform Assessment:} 
\begin{itemize}
\setlength{\itemsep}{0pt}
\setlength{\parskip}{0pt}
\setlength{\parsep}{0pt}  
\item Does this platform follow any recognized standardised?
\item How matured and well used are the standards?
 
\end{itemize}

\noindent \textbf{Application Assessment:}

\begin{itemize}
\setlength{\itemsep}{0pt}
\setlength{\parskip}{0pt}
\setlength{\parsep}{0pt}  
\item Does this application follow any recognized standardised? 
\item How matured and well used are the standards?
\begin{itemize}
\item Infrastructure (e.g.,: 6LowPAN, IPv4/IPv6, RPL)
\item Identification (e.g.,: EPC, uCode, IPv6, URIs)
\item Comms / Transport (e.g.,: Wifi, Bluetooth, LPWAN)
\item Discovery (e.g.,: Physical Web, mDNS, DNS-SD, HyperCat)
\item Data Protocols (e.g.,: MQTT, CoAP, AMQP, Websocket, Node)
\item Device Management (e.g.,: TR-069, OMA-DM)
\item Semantic (e.g.,: JSON-LD, Web Thing Model)
\item Multi-layer Frameworks (e.g.,: Alljoyn, IoTivity, Weave, Homekit)

\end{itemize}

\end{itemize}
\end{monsterboxnobg}


\begin{paperbox}{30. Compliance \hfill \tikzcircledemo{4pt}}
Based on the country and region, there are number of policies, laws and regulations that need to be adhered to. It is important for IoT applications to respect guidelines. Some  regulatory efforts are  ISO 29100 Privacy framework, OECD privacy principles, and European Commission Protection of personal data.
\end{paperbox}
\begin{monsterboxnobg}{}
\noindent \textbf{Platform Assessment:} 
\begin{itemize}
\setlength{\itemsep}{0pt}
\setlength{\parskip}{0pt}
\setlength{\parsep}{0pt}  
\item Where is this platform is going to be in operation?
\item Based on the above location, what are the laws and guidelines that the platform need to be adhered to?
\item Does the platform compliance  to such laws and regulations?
\item Does th platform allows developers to pick and choose a particular region with respected to the privacy laws and automatically configures itself to compliance with the selected regions?
\item Does the platform aware of privacy laws and regulations in different countries and regions?
 
\end{itemize}

\noindent \textbf{Application Assessment:}

\begin{itemize}
\setlength{\itemsep}{0pt}
\setlength{\parskip}{0pt}
\setlength{\parsep}{0pt}  
\item Where is this application is going to be in operation?
\item Based on the above location, what are the laws and guidelines that the application need to be adhered to?
\item Does the application compliance s to such laws and regulations?
\end{itemize}
\end{monsterboxnobg}

\noindent \textbf{References}
\begin{itemize}
\item Perera, C.; McCormick, C.; Bandara, A. K.; Price, B. A. \& Nuseibeh, B. \textit{Privacy-by-Design Framework for Assessing Internet of Things Applications and Platforms}, Proceedings of the 6th International Conference on the Internet of Things, ACM, 2016, 83-92

\item Perera, C.; Bandara, A. K.; Price, B. \& Nuseibeh, B. Designing \textit{Privacy-aware Internet of Things Applications}, arXiv preprint arXiv:1703.03892, 2017
\end{itemize}

%

%

\end{document}